\documentclass[12pt,preprint]{aastex}
\usepackage{graphicx}
\usepackage{longtable}
\usepackage{multirow}
\usepackage{url}
\usepackage{lipsum}
\def\keV{\,{\rm keV}}

\begin{document}

\title{The formation rate of short gamma-ray bursts and gravitational waves}

\author{G. Q. Zhang\altaffilmark{1} and F. Y. Wang\altaffilmark{1,2}{*}}
\affil{
$^1$ School of Astronomy and Space Science, Nanjing University, Nanjing 210093, China\\
$^2$ Key Laboratory of Modern Astronomy and Astrophysics (Nanjing University), Ministry of Education, Nanjing 210093, China\\
} \email{{*}fayinwang@nju.edu.cn}

\begin{abstract}
In this paper, we study the luminosity function and formation rate
of short gamma-ray bursts (sGRBs). First, we derive the $E_p-L_p$
correlation using 16 sGRBs with redshift measurements and determine
the pseudo redshifts of 284 \textit{Fermi} sGRBs. Then, we use the
Lynden-Bell c$^-$ method to study the luminosity function and
formation rate of sGRBs without any assumptions. A strong evolution
of luminosity $L(z)\propto (1+z)^{4.47}$ is found. After removing
this evolution, the luminosity function is $ \Psi (L) \propto L_0 ^
{- 0.29 \pm 0.01} $ for dim sGRBs and $ \psi (L) \propto L_0 ^ {-
1.07 \pm 0.01} $ for bright sGRBs, with the break point $8.26 \times
10^{50} $ erg s$^{-1}$. We also find that the formation rate
decreases rapidly at $z<1.0$, which is different from previous
works. The local formation rate of sGRBs is 7.53 events Gpc$^{-3}$
yr$^{-1}$. Considering the beaming effect, the local formation rate of
sGRBs including off-axis sGRBs is $ 203.31^{+1152.09}_{-135.54} $
events Gpc$^{-3}$ yr$^{-1}$. We also estimate that the event rate of
sGRBs detected by the advanced LIGO and Virgo is $0.85^{+4.82}_{-0.56} $
events yr$^{-1}$ for an NS-NS binary.
\end{abstract}


\section{Introduction}\label{introduction}
Gamma-ray bursts (GRBs) are the most violent explosions in the
universe \citep{Berger14,Kumar15,Wang15}. They can be divided into two groups,
short GRBs (sGRBs) and long GRBs, with a separation at about 2
s \citep{Kouveliotou93}. The uniform distribution on the
sky, the $\log N -\log S$ correlation, and
the discovery of the afterglows of GRB 050509B
\citep{2005Natur.437..851G},  GRB 050709
\citep{2005Natur.437..855V}, and GRB 050724 \citep{2005Natur.438..994B}
demonstrate that sGRBs have a cosmological origin. Therefore, sGRBs
are powerful tools with which to study the universe.

Although the study of sGRBs developed rapidly, the central engine of
sGRBs is still under debate \citep{2011ApJ...727..109V}.
The most popular model is the merge of the compact object binary
\citep{1989Natur.340..126E, 1992ApJ...395L..83N}, such as binary
neutron stars (NSs) or black hole (BH)-neutron star binaries. The
location of sGRBs \citep{2013ApJ...776...18F} and the presence of
kilonova emission \citep{2013ApJ...774L..23B,2013Natur.500..547T}
provide evidence for this model. One important prediction of this
model is that sGRBs should be accompanied with gravitational wave
radiation \citep{1992Sci...256..325A,1992ApJ...395L..83N}. Recently,
advanced LIGO detected some
gravitational wave events, GW 150914 \citep{2016PhRvL.116f1102A},
GW 151216 \citep{2016PhRvL.116x1103A}, GW 170104
\citep{2017PhRvL.118v1101A}, and GW 170814
\citep{2017PhRvL.119n1101A}. Notably, GW 170817, which is associated
with GRB 170817A, is believed to originate from the merge of binary neutron
stars, which  provides strong evidence for this model
\citep{2017PhRvL.119p1101A}.

The formation rate of long GRBs has been extensively explored
\citep{Kistler08,Wang09,Wanderman10,Coward13}. Although the number of
observed sGRBs is increasing, only a small fraction of sGRBs
have redshift measurements. Therefore, it is very difficult to
estimate the sGRB formation rate. The local formation rate, estimated
by previous works, ranges from 0.1 events Gpc$^{-3}$ yr$^{-1}$ to 400
events Gpc$^{-3}$
yr$^{-1}$\citep{2005A&A...435..421G,2006A&A...453..823G,2006ApJ...650..281N,2009A&A...498..329G,2014MNRAS.437..649S,2015ApJ...812...33S,2017arXiv171005620P}.
These results are mainly determined by fitting the peak flux
distribution with a given function form. So the derived formation
rate of sGRBs is model dependent. Another challenge to estimate the
formation rate is the selection effects for sGRBs. The most
important selection effect is the flux limit of satellites, thus, the
observed data may ignore some dim sGRBs. The Lynden-Bell c$^-$
method can deal with a flux-limit sample, which is proposed by
\citet{1971MNRAS.155...95L}. This method has been used in long GRBs
\citep{2002ApJ...574..554L,2004ApJ...609..935Y,2006ApJ...642..371K,2012MNRAS.423.2627W,2015ApJ...806...44P,2015ApJS..218...13Y}.
Using the $E_p-L_p$ correlation and the Lynden-Bell c$^-$ method,
\citet{2014ApJ...789...65Y} found that the formation rate of sGRBs
follows SFRs at $z<4.0$ with local rate $0.63^{+0.31}_{-0.39}$
events Gpc$^{-3}$ yr$^{-1}$. Besides,
\citet{2017PhRvL.119p1101A} obtained the local formation rate of
binary neutron stars, which is $1540^{+3200}_{-1200}$ Gpc$^{-3}$ yr$^{-1}$
based on the detection of GW 170817 at about 40 Mpc. Several selection effects, such as Malmquist bias, redshift desert, and the fraction of afterglows missing because of host
galaxy dust extinction, are considered for \textit{Swift} long GRBs \citep{Coward13}.

In this paper, the $E_p-L_p$ correlation is obtained with 16 sGRBs.
Using this correlation, we get the pseudo redshifts of 284 \textit{Fermi} sGRBs.
Then we derive the luminosity function and formation rate of sGRBs
using the Lynden-Bell c$^-$ method. Last, we estimate the number of
gravitational wave events associated with sGRBs. This paper is
organized as follows. In the next section, we derive the $E_p-L_p$
correlation and the pseudo redshifts of \textit{Fermi}/GBM sGRBs. In section 3,
the luminosity function and formation rate of sGRBs are determined
using Lynden-Bell c$^-$ method. In section 4, we test our results
with Monte Carlo simulations. Finally, we give conclusions and
discussion in section 5. Throughout this paper, we adopt the flat
$\Lambda$CDM model with $\Omega_m = 0.27$ and $H_0 = 70$ km s$^{-1}$
Mpc$^{-1}$.

\section{sGRB Sample}
\label{sec:Sample}

The number of sGRBs with measured redshifts is very small. In order
to constrain the luminosity function and formation rate, we
derive the pseudo redshifts of sGRBs using the
correlation between the peak energy $E_p$ and the peak
luminosity $L_p$ \citep{2013MNRAS.431.1398T}. In this section, we fit the $E_p-L_p$ correlation with
more sGRBs.

We collect 16 sGRBs with measured redshifts.
In Table 1, we list the properties of these 16 sGRBs, including name,
duration in the rest-frame $T_{90}^{rest} = T_{90} / (1 +z)$,
measured redshift, spectral peak energy $E_p$ and peak luminosity
$L_p$ in 64 ms of the observer-frame time bin. It should be noted
that the first eight sGRBs are consistent with those of
\citet{2013MNRAS.431.1398T}. The Band function
\citep{1993ApJ...413..281B}
\begin{equation}
  \label{eq:1}
  f(E) = \left\{ \begin{tabular}{ll}
$ A(\frac{E}{100 \keV})^\alpha \times \exp(- \frac{(2 + \alpha)E}{E_p}) $  $E < \frac{(\alpha - \beta)E_p}{2 + \alpha}. $\\
$ A(\frac{E}{100 \keV})^\beta \times \exp^{\beta -
\alpha}(\frac{(\alpha - \beta)E_p}{(2 + \alpha) 100 \keV})^{\alpha -
\beta} $  $E \geqslant \frac{(\alpha - \beta)E_p}{2 + \alpha}, $
\end{tabular} \right.
\end{equation}
is used to fit the spectra of sGRBs.
The peak luminosity is calculated in the 1-10$^5$
keV energy range. The photon indices are set as $\alpha = -1$ and $\beta = -2.25$
for those that lack of observational constraints \citep{Schaefer07,Wang11}. For sGRBs observed by
\textit{Swift}, the 64 ms peak flux is estimated by correcting the 1 s peak
flux with the ratio of the 64 ms and 1024 ms peak counts in
the 64 ms binned light curve provided by \textit{Swift} \citep{2015MNRAS.448.3026W}.

\begin{figure}
    \centering
    \includegraphics[width=0.8\linewidth]{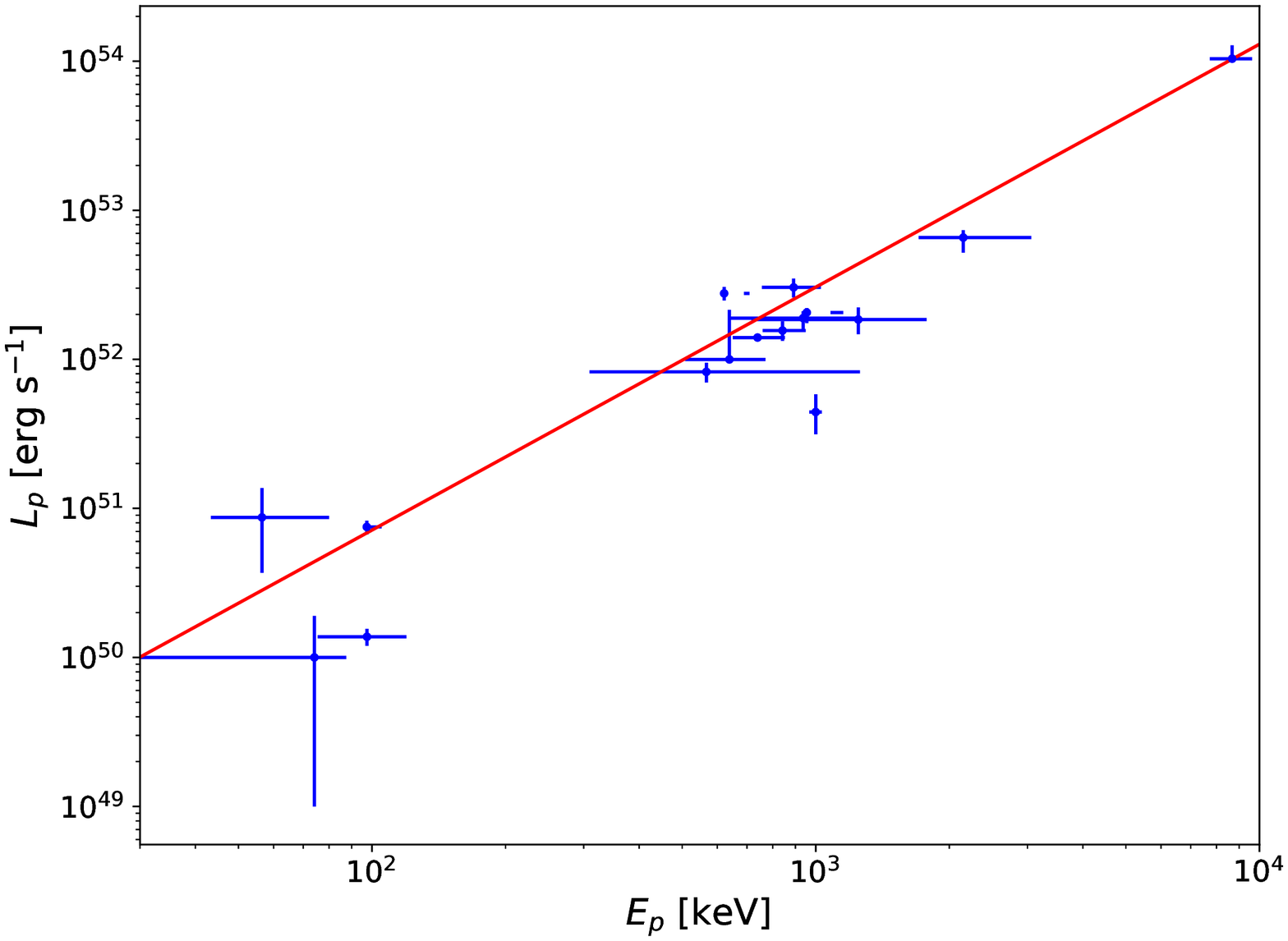}
    \caption{$E_p-L_p$ correlation for sGRBs. The red line is the best fit.}
    \label{fig:fit}
\end{figure}

We fit the $E_p-L_p$ correlation with the linear form
\begin{equation}
  \label{eq:2}
  L_p = a (\frac{E_p}{100 \keV})^b.
\end{equation}
The best fit of this correlation is given by
\begin{equation}
  \label{eq:3}
  L_p = (7.15 \pm 0.49) \times 10^{50} (\frac{E_p}{100 \keV})^{1.63 \pm
  0.03},
\end{equation}
which is consistent with the correlation obtained by
\citet{2013MNRAS.431.1398T}. Their coefficients are
$a = 7.5 \times 10^{50}$ and $b = 1.59$. \cite{2017arXiv171005620P} also fit
this correlation and found $a = 5.71 \times 10^{50},
b = 1.73$. These results are consistent with each other.
Figure \ref{fig:fit} shows the
$E_p-L_p$ correlation. The red line is the best-fitting result. We
can rewrite the Equation (\ref{eq:3}) as
\begin{equation}
  \label{eq:5}
   \frac{d_l^2}{(1 + z)^{1.63}} = \frac{7.15 \times 10^{50}}{4\pi F_p}(\frac{E_p}{100
   \keV})^{1.63},
\end{equation}
where $d_l$ is the luminosity distance, $F_p$ is the peak flux at
64ms time intervals, and $E_p$ is the peak energy. The pseudo
redshifts can be derived using the observed peak energy $E_p$ and 64
ms peak flux $F_p$. The sGRB 170817A associated with
GW 170817 is detected by \textit{Fermi}/GBM. It is interesting to test whether
this sGRB follows the $E_p-L_p$ correlation. From GBM observations,
the peak luminosity is $L_{p,off}=(1.6\pm 0.6)\times 10^{47}$ erg s$^{-1}$
and $E_{p,off}=200$ keV \citep{Fermi17}. Obviously, GRB 170817A
deviates the best fit of the $E_p-L_p$ correlation, which is derived
from on-axis sGRB values. However, note that $L_{p,off}$
and $E_{p,off}$ are off-axis values. If we consider uniform top-hat
jets with a sharp edge \citep{Rhoads97} widely used to explain GRB
properties, the on-axis values are
$L_{p,on}=L_{p,off}[1+\Gamma^2(\zeta-\theta_j)^2]^3$ and
$E_{p,on}=E_{p,off}[1+\Gamma^2(\zeta-\theta_j)^2]$
\citep{Granot02,Fermi17}, where $\Gamma$ is the Lorentz factor of
the jet, $\zeta$ is the viewing angle, and $\theta_j$ is the initial jet
opening angle. The GW fitting parameters suggest that the viewing angle
is less than 56$^{\circ}$ \citep{2017PhRvL.119p1101A}. However, the
values of $\Gamma$ and $\theta_j$ are unclear \citep{Evans17,Zou17}.
If Lorentz factor $\Gamma=50$ and $\zeta-\theta_j=11^{\circ}$ are
assumed, GRB 170817A follows the $E_p-L_p$ correlation.

We select 284 sGRBs observed by \textit{Fermi}/GBM
\citep{2014ApJS..211...12G,2014ApJS..211...13V,2016ApJS..223...28N}.
The pseudo
redshifts of these sGRBs are obtained through equation (\ref{eq:5}).
In table \ref{tab:fermi}, we list the name of sGRB, the rest-frame
duration $T_{90}$, the pseudo redshift $z$, the peak flux $F_p$ at 64 ms
time intervals, the peak energy $E_p$ , the photon indices $\alpha,
\beta$ and the peak luminosity $L_p$ within energy range 1-10$^5$
keV. The photon indices are set as $\alpha=-1, \beta=-2.25 $ for
sGRBs without observational constraints.

\begin{figure}
    \centering
    \includegraphics[width=0.8\linewidth]{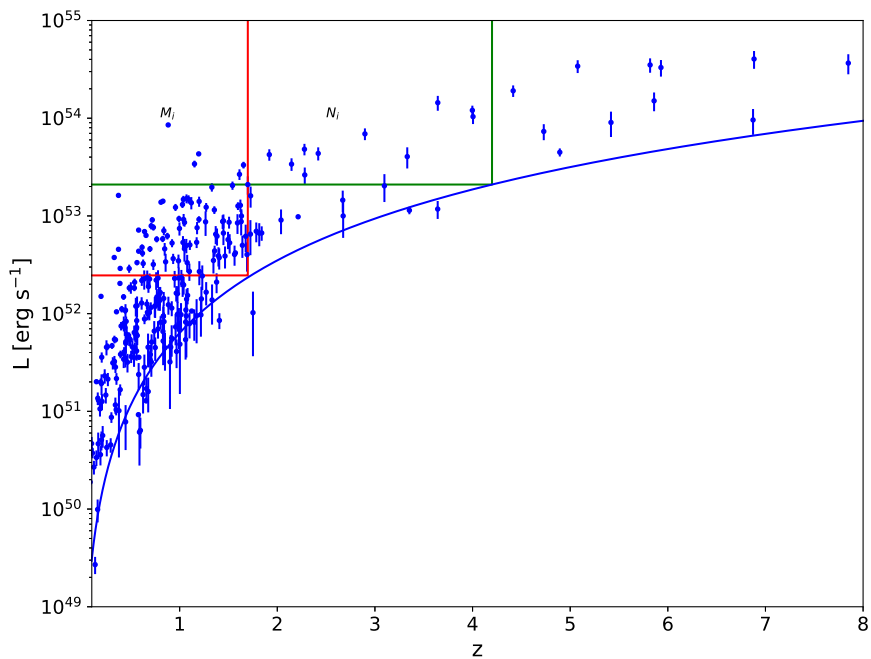}
    \caption{Redshift distribution of sGRBs estimated by the $E_p-L_p$
        correlation. The blue line is the flux limit of $ 2.3 $  photons
        cm$^{-2}$ s$^{-1}$}
    \label{fig:fermilz}
\end{figure}

In order to study the luminosity function, we introduce the
flux limit of the \textit{Fermi}/GBM. According to
\citet{2003ApJ...588..945B}, the GBM flux limit is weakly dependent on
peak energy $E_p$. Therefore, we set the flux limit as a
constant, which has been widely used in the literature.
\citet{2003ApJ...588..945B} calculated the flux limit for accumulation
time $\Delta t=1$ s. After converting it to $\Delta t$ = 64 ms, we find that the flux limit
is about 2.0 photons cm$^{-2}$ s$^{-1}$. Besides, the flux
distribution of sGRBs, which were observed by \textit{Fermi}/GBM indicated that the
flux limit on a 64 ms timescale is 2.3 photons cm$^{-2}$ s$^{-1}$
\citep{2016ApJS..223...28N}. This value is also used in
\citet{2015ApJ...809...53C}. Therefore, we set the photon flux
limit as 2.3 photons cm$^{-2}$ s$^{-1}$. The distribution of the
pseudo redshifts and 64 ms luminosity is shown in Figure
\ref{fig:fermilz}. The blue line is the flux limit $F_{limit} = 2.3$
photons cm$^{-2}$ s$^{-1}$. We remove the sGRBs with
$z > 3$, because the maximum redshift observed for sGRBs is 2.609.
Hereafter, we use 239 sGRBs that are brighter than the flux limit for further
analysis.

\section{Luminosity function and formation rate}
\label{sec: lynden}

\subsection{Lynden-Bell $c^{-}$ Method}
\label{subsec:method} Many previous works fitted the luminosity
function with a given function form. These results have a strong
dependence on the function form. In this paper, we use the
Lynden-Bell c$^-$ method to derive the luminosity function and
the formation rate of sGRBs. The Lynden-Bell $c^{-}$ method is an efficient
method to determine the distribution of the redshift and the luminosity
function of astronomical objects with a truncated sample.
\citet{1971MNRAS.155...95L} developed this method and derived the
luminosity function and density evolution for quasars. This method
has been used to study pulsars \citep{2016Ap&SS.361..138D}, long GRBs
\citep{2002ApJ...574..554L,2004ApJ...609..935Y,2015ApJ...806...44P,2015ApJS..218...13Y,
Tsvetkova17}
and sGRBs \citep{2014ApJ...789...65Y}.

If the luminosities and redshifts of sGRBs are independent, the
distribution of luminosity and redshift should be $\Psi(L,z) =
\psi(L)\rho(z)$, where $ \psi(L) $ is the luminosity function and $
\rho(z) $ represents the formation rate of sGRBs. However, there
exists a significant degeneracy between luminosity and redshift
\citep{2002ApJ...574..554L}. Therefore, $\Psi(L,z)$
should be written as $ \Psi (L,z) = \rho (z)
\psi (L/g(z)) / g(z)$, where $ g(z) $ is the correlation between the
luminosity and redshift. The luminosity at redshift $ z = 0 $ is
$L_0 = L / g(z)$. The goal of our analysis is to obtain the
formation rate $\rho (z)$ and the local luminosity function $\psi
(L/g(z))$.

The first step is to remove the effect of luminosity evolution. We
take the evolution function form as $g(z) = (1+z)^k$, which has been
used in previous works \citep{2002ApJ...574..554L,
2014ApJ...789...65Y,2015ApJS..218...13Y}. In order to determine the
value of $k$, we use the $\tau$ statistical method
\citep{1992ApJ...399..345E}. For each point $(L_i, z_i)$, we can
define the associated set $J_i$ as
\begin{equation}
  \label{eq:6}
   J_i = \{j| L_j \geqslant L_i, z_j \leqslant z_i^{\rm{max}}\},
\end{equation}
where $L_i$ is the luminosity of $i$th sGRB and $z_i^{\rm{max}}$ is the
maximum redshift at which the sGRBs with the luminosity $L_i$ can be
detected by satellite. We plot this region as a green rectangle in
figure \ref{fig:fermilz}. We define the number of sGRBs in this
region as $n_i$ and the number of sGRBs with redshifts less than or
equal to $z_i$ as $R_i$. The expected mean and the variance of $R_i$
should be $E_i = \frac{n_i + 1}{2}$ and $V_i = \frac{n_i^2 -
1}{12}$, respectively.

\begin{figure}
    \centering
    \includegraphics[width=0.8\linewidth]{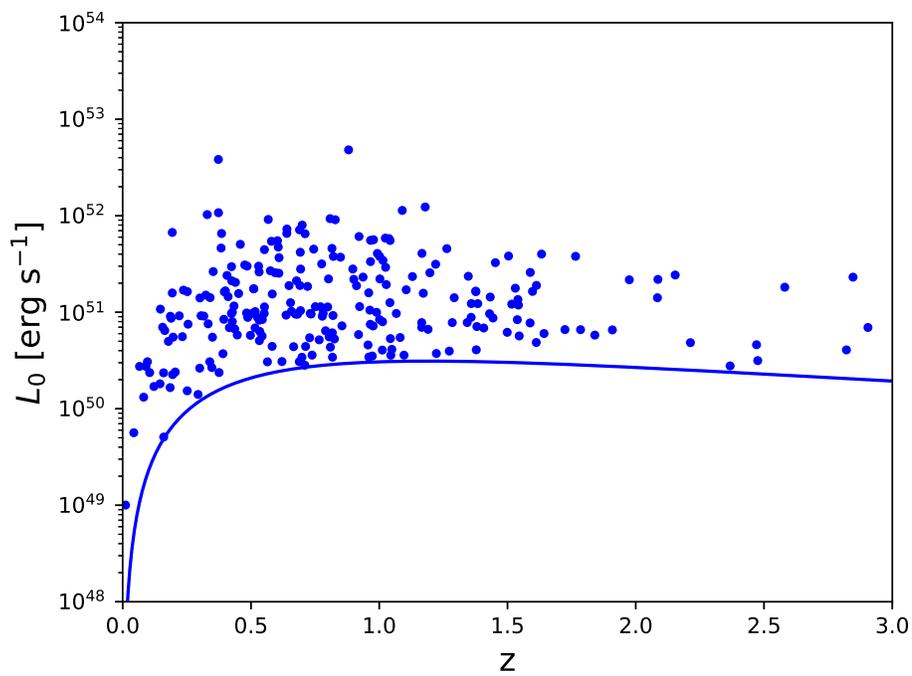}
    \caption{Non-evolving luminosity $L_0 = L/(1+z)^{4.47}$ of 284 sGRBs. The blue line is the flux limit.}
    \label{fig:fermil0-a}
\end{figure}

The statistic $\tau$ to test the dependence between
luminosity and redshift is
\begin{equation}
  \label{eq:7}
   \tau =\frac{ \sum_i (R_i - E_i)}{\sqrt{ \sum_i V_i}}.
\end{equation}
If luminosity and redshift are independent, $R_i$ should be
uniformly distributed between 1 and $n_i$. Therefore, $\tau$ should
be zero.
We change the value of $k$ until the test statistic $\tau$ is zero.
Finally, we find that the best fitting is $k = 4.47^{+0.47}_{-0.29}$, which is
similar to \citet{2017arXiv171005620P}. He obtained $k = 4.269 \pm 0.134$ for
\textit{Fermi} sGRBs. We
show the distribution of non-evolving luminosity and redshift in
Figure \ref{fig:fermil0-a}.

\subsection{Luminosity Function}
\label{subsec: luminosity} 
\begin{figure}
    \centering
    \includegraphics[width=0.8\linewidth]{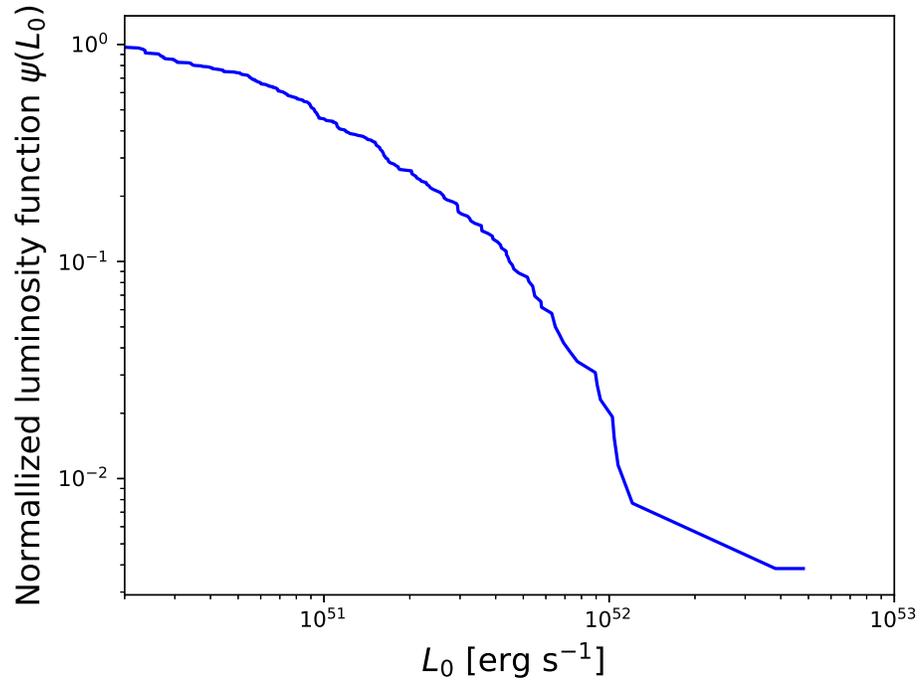}
    \caption{Cumulative luminosity function of $L_0$, which is normalized to unity
        at the first point.}
    \label{fig:fermipsil}
\end{figure}

After removing the effect of luminosity
evolution through $L_0 = L / (1+z)^k$, the cumulative luminosity function
can be derived with a nonparametric method from the following
equation \citep{1971MNRAS.155...95L,1992ApJ...399..345E}
\begin{equation}
  \label{eq:8}
  \psi(L_{0i}) = \prod_{j < i} (1 + \frac{1}{N_j}),
\end{equation}
where $j < i$ means that the $j$th sGRB has a larger luminosity than
$i$th sGRB. The cumulative luminosity function is shown in Figure
\ref{fig:fermipsil}. We use a broken power-law form to fit the
luminosity function. The best fit is given by
\begin{equation}
  \label{eq:9}
  \psi(L_0) \propto\left\{\begin{array}{ll}
                      L_0^{-0.29 \pm 0.01} & L_0 < L_0^b \\
                      L_0^{-1.07 \pm 0.01} & L_0 > L_0^b \\
                    \end{array}
                    \right.
\end{equation}
with the break luminosity $L_0^b = 8.26 \times 10^{50}$ erg
s$^{-1}$. Our result is marginally consistent with
\citet{2014ApJ...789...65Y} for bright sGRBs. They get $\psi(L_0)
\propto L_0^{-0.84}$ between the luminosity $10^{51}$ and $10^{53}$
erg s$^{-1}$. It should be noted that this is the luminosity
function at $z = 0$. The luminosity function at redshift $z$ is
$\psi(L_0)(1+z)^{4.47}$.

\subsection{sGRB Formation Rate}
\label{subsec:formation} To obtain the formation rate of sGRBs, we
define $J_i'$ as
\begin{equation}
  \label{eq:10}
  J'_i = \{j| L_j  > L_i^{\rm{lim}}, z_j < z_i\},
\end{equation}
where $z_i$ is the redshift of $i$th sGRB, and $L_i^{\rm{lim}}$ is the
minimum luminosity, which can be observed at redshift $z_i$. The
number of sGRBs in this region is $M_i$. Similar to deriving the
luminosity function, we can obtain the cumulative redshift
distribution as
\begin{equation}
  \label{eq:11}
  \phi(z_i) = \prod_{j < i} (1 + \frac{1}{M_j}).
\end{equation}
The formation rate of sGRBs can be derived from
\begin{equation}
  \label{eq:12}
  \rho(z) = \frac{d\phi(z)}{dz}(1 + z)(\frac{dV(z)}{dz})^{-1},
\end{equation}
where the factor $(1+z)$ is due to the cosmological time dilation.
The differential comoving volume $dV(z)/dz$ is
\begin{equation}
  \frac{dV(z)}{dz} = 4\pi (\frac{c}{H_0})^3(\int^z_0 \frac{dz}{\sqrt{1 - \Omega_m + \Omega_m(1 + z)^3}})^2  \times \frac{1}{\sqrt{1 - \Omega_m + \Omega_m(1 + z)^3}}. \\
\end{equation}

\begin{figure}
    \centering
    \includegraphics[width=0.8\linewidth]{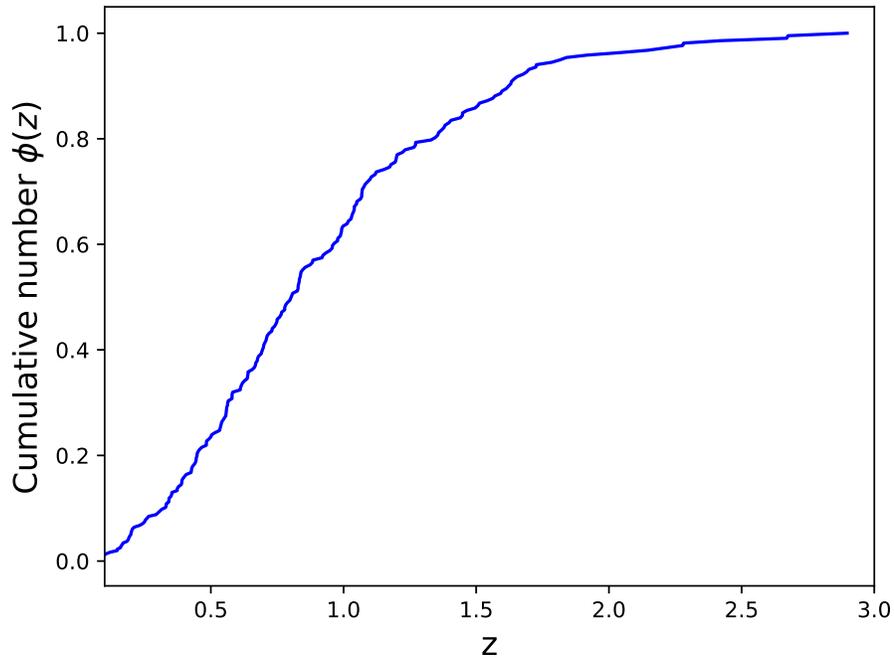}
    \caption{Cumulative redshift distribution of sGRBs.}
    \label{fig:fermiphiz}
\end{figure}

\begin{figure}
    \centering
    \includegraphics[width=0.8\linewidth]{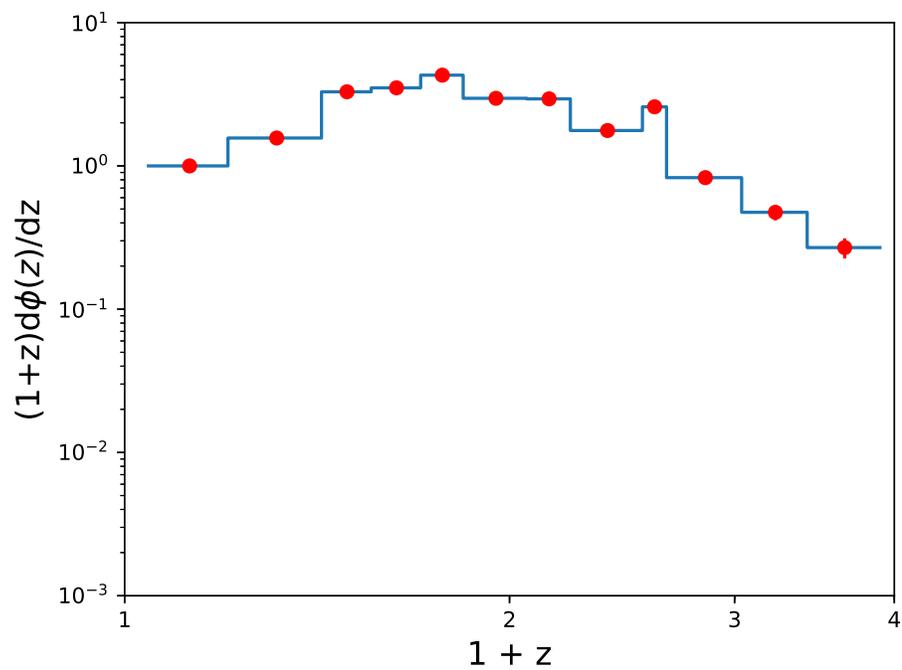}
    \caption{Evolution of $(1+z)\frac{d\phi(z)}{dz}$ as a function of redshift $z$ with 1$\sigma$ errors, which is normalized to unity at the first point.}
    \label{fig:figure1}
\end{figure}

\begin{figure}
    \centering
    \includegraphics[width=0.8\linewidth]{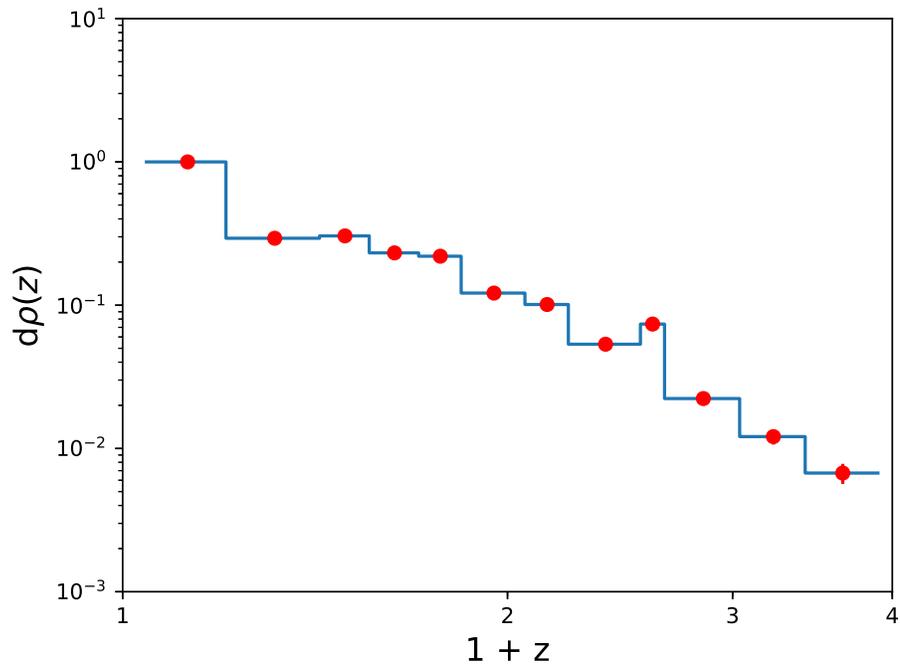}
    \caption{Comoving formation rate of sGRBs, which is normalized to unity at the first point.}
    \label{fig:fermirou}
\end{figure}

The cumulative redshift distribution is plotted in Figure
\ref{fig:fermiphiz}.  We also show $ (1 + z) \frac{d\phi(z)}{dz}$ in
Figure \ref{fig:figure1}. From this figure, it is obvious that
$(1+z)d\phi(z)/dz$ increases quickly at $z < 0.8$, remains constant
at $0.8 < z < 1.5$, and then decrease quickly at $z
> 1.5$. This profile is similar as the formation rate derived by \citet{2014ApJ...789...65Y}.
Figure \ref{fig:fermirou} gives the formation rate of sGRBs. The
best fitting of $\rho(z)$ is
\begin{equation}
  \label{eq:14}
  \rho(z) \propto \left\{
    \begin{array}{ll}
      (1+z)^{-3.08 \pm 0.06} & z < 1.60 \\
      (1+z)^{-4.98 \pm 0.03} & z \geq 1.60
    \end{array}
  \right.
\end{equation}
The formation rate $\rho(z)$ decreases at all redshift ranges, which
is dramatically different from \citet{2014ApJ...789...65Y}. Because the
result of \citet{2014ApJ...789...65Y} is similar to Figure
\ref{fig:figure1}, they may omit the differential comoving volume
$dV(z)/dz$ term in their calculation, which has been discussed in
\citet{2015ApJS..218...13Y}. The local event rate is 7.53 Gpc$^{-3}
$ yr$^{-1}$. \citet{2015ApJ...815..102F} also found that the local
formation rate is 10 Gpc$^{3}$ yr$^{-1}$, which is consistent with
our results.

\section{Testing with the Monte Carlo Simulation}
\label{sec:label} In this section, we use Monte Carlo simulations to
test our results. We simulate a set of points $(L_0, z)$, which
satisfies Equations (\ref{eq:9}) and (\ref{eq:14}). The
luminosity function of Equation (\ref{eq:9}) is at redshift $z = 0$.
So we transform the luminosity $L_0$  to $L = L_0(1+z)^{4.47}$. Then
we obtain a set of points $(L,z)$, which is similar to observed
data. We simulate 200,000 points and divide them into 200 groups. In
each group, 200 points are selected as pseudo data points. We
perform the same analysis as above to obtain the luminosity function
and formation rate of sGRBs.

\begin{figure*}
    \centering
    \includegraphics[width=1.0\linewidth]{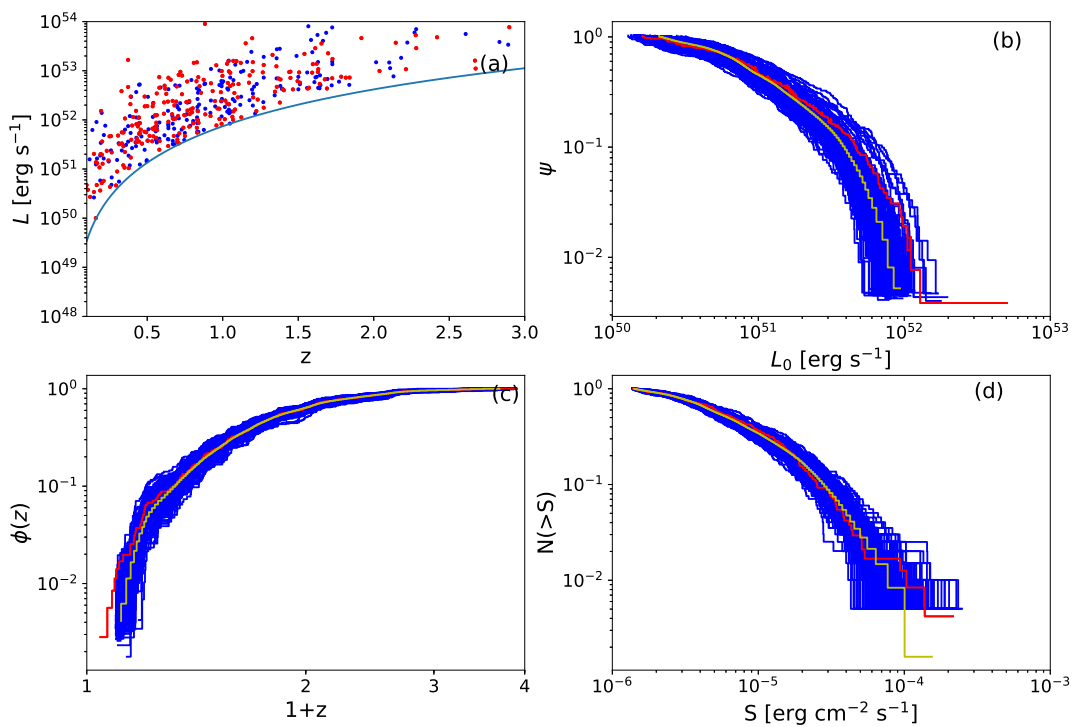}
    \caption{Comparison of simulated data (blue) and observed data (red). These
        panels show luminosity-redshift distribution (panel (a)), cumulative luminosity
        function (panel (b)), cumulative redshift distribution (panel (c)) and $\log N-\log S$ distribution (panel (d)), respectively.
        The yellow lines are the mean of simulated data. }
    \label{fig:mcmc}
\end{figure*}

Our results are shown in Figure \ref{fig:mcmc}. In panel (a), the
luminosity-redshift distribution is shown. We select one sGRB from
each group and get 200 pseudo sGRBs to compare with the observed
data. The blue points and red points
represent the simulated data and observed data, respectively. The
solid line is the flux limit 2.3 photons cm$^{-2}$
s$^{-1}$. It is obvious that the simulated data and the observed data
have a similar distribution. In panels (b),(c), and (d), the blue
lines are the cumulative luminosity function, cumulative redshift
distribution, and $\log N - \log S$ distribution derived from the
simulated data. The red lines and yellow lines are observed data and
the mean of the simulated data, respectively. We perform the
Kolmogorov-Smirnov test between observed data and the mean
distribution of simulated data. The $p$ value for panels (b), (c),
and (d) are 0.29, 0.99, and 0.94, respectively. Meanwhile, the
distributions of the observed data lie in the region of the
simulated data. Therefore, the cumulative luminosity function and
formation rate are reliable.

\section{Conclusions and Discussion}
\label{sec:conclusions} In this paper, we first use 16 sGRBs with
measured redshifts to fit the $E_p-L_p$ correlation. Using this
correlation, we obtain the pseudo redshifts of 284 sGRBs observed by
\textit{Fermi}/GBM, which are listed in Table \ref{tab:fermi}. Then, the
Lynden-Bell $c^{-}$ method is used to study the luminosity function
and the formation rate of sGRBs. The effect of luminosity evolution is
removed by $L_0 = L / (1 + z)^k$, where $k = 4.47^{+0.47}_{-0.29}$.
After removing the effect of luminosity evolution, we derive the
cumulative luminosity function. The result is shown in Figure
\ref{fig:fermipsil}, which can be fitted with a broken power law as
$\psi(L_0) \propto L_0^{-0.29 \pm 0.01}$ for $L_0 < L_0^b$ and
$\psi(L_0) \propto L_0^{-1.07 \pm 0.01}$ for $L_0 > L_0^b$, with
$L_0^b=8.26 \times 10^{50}$  erg s$^{-1}$. \citet{2015MNRAS.448.3026W} found the
luminosity function with power-law indices $-0.94$ for dim bursts
and $-2.0$ for luminous bursts. The break luminosity is $2 \times
10^{52}$ erg s$^{-1}$. These indices and the break point are much larger
than our results. The reason is that they do not consider the effect
of luminosity evolution. Considering the luminosity evolution,
\citet{2014ApJ...789...65Y} determined the index
$-0.84^{+0.07}_{-0.09}$ for bright sGRBs, which is consistent with
our result.

In Figure \ref{fig:figure1}, we show the evolution of
$(1+z)\frac{d\phi(z)}{dz}$, which increases quickly at $z < 0.8$,
remains approximately constant at $ 0.8 < z < 1.5$, and decreases
rapidly at $ z > 1.5$. Figure \ref{fig:fermirou} shows the formation
rate of sGRBs. We find that the formation rate is decreasing
quickly. The best fit is $ \rho(z) \propto(1 + z)^{-3.08} $ for $z < 1.60$ and
$\rho(z) \propto(1+z)^{-4.98} $ for $z \geq 1.60$. Obviously, the formation rate is in contrast to
previous estimations by
\citet{2015MNRAS.448.3026W}, \citet{2016A&A...594A..84G}, and \citet{2014ApJ...789...65Y}.
By assuming the formation rate of sGRBs has a time delay to the
star formation rate, \citet{2015MNRAS.448.3026W} and \citet{2016A&A...594A..84G}
found that the formation rate is increasing at $z < 1.5$ and decreasing
at $ z> 1.5$. The Lynden-Bell c$^-$ method is also used by
\citet{2014ApJ...789...65Y} to study the formation rate of sGRBs.
They found the formation rate increases at $z < 0.6$ and remains
constant at $0.6 < z < 2$, which is similar to the evolution of $(1+z)\frac{d\phi(z)}{dz}$ shown in figure
\ref{fig:figure1}. Therefore, they may omit the differential
comoving volume term.

The local formation rate of sGRBs is 7.53 Gpc$^{-3}$ yr$^{-1} $,
which is consistent with that of \citet{2015ApJ...815..102F}. If
only the GRB 170817A that occurred at about 40 Mpc is considered, the
local rate of sGRBs is $\sim 463$ Gpc$^{3}$ yr$^{-1}$. \citet{2017arXiv171005851Z}
obtained that the event rate is $190^{+440}_{-160}$ Gpc$^3$ yr$^{-1}$,
which is similar with our result. Some
estimations are lower than our result, for example, $0.51
^{+0.36}_{-0.19}$ Gpc$^{-3}$ yr$^{-1}$ \citep{2004JCAP...06..007A} and
$0.63^{+0.31}_{-0.39}$ Gpc$^{-3}$ yr$^{-1}$
\citep{2014ApJ...789...65Y}. Besides, \citet{2006ApJ...650..281N}
and \citet{2006A&A...453..823G} obtained the formation rate as $40
\pm 12$ Gpc$^{-3}$ yr$^{-1}$  and $30^{+50}_{-20}$ Gpc$^{-3}$
yr$^{-1}$ , respectively, which are larger than our result.

If we set the beaming factor as $f_b^{-1} = 27 ^{+158}
_{-18}$\citep{2015ApJ...815..102F}, the local event rate of sGRBs
including the off-axis ones is $\rho_{0,all} = 203.31
^{+1152.09}_{-135.54}$  Gpc$^{-3}$ yr$^{-1}$. At present, the
horizon of aLIGO and Virgo for the merger of NS-NS is 80-120 Mpc
\citep{2016LRR....19....1A}. If we suppose that the sGRBs arise from the
mergers of NS-NS binaries, the event rate of the gravitational wave is
$0.85^{+4.82}_{-0.57} $ events yr$^{-1}$. The horizon for NS-NS will
increase to 200 Mpc in 2019 \citep{2016LRR....19....1A}. Then aLIGO
will detect $ 6.81^{+38.59}_{-4.53} $ events per year. If we assume
that the mass of a black hole is 5 $M_\odot$, the range of BH-NS is
approximately a factor of 1.6 larger than NS-NS range. Thus it is
excepted to detect $3.48 ^{+19.75}_{-3.48}$ events every year for
now and $27.89^{+158.05}_{-18.59} $ events every year in 2022.

\section*{Acknowledgments} We thank the anonymous referee for constructive comments.
We acknowledge the use of public data from
\textit{Fermi}. We thank Hai Yu for helpful discussions. This work is
supported by the National Basic Research Program of China (973
Program, grant No. 2014CB845800), the National Natural Science
Foundation of China (grants 11422325 and 11373022), and the
Excellent Youth Foundation of Jiangsu Province (BK20140016).

\newpage

\newpage

\newpage

\newpage

\newpage

\newpage

\newpage

\newpage

\newpage

\begin{deluxetable}{rrrrrr}
    \tabletypesize{\footnotesize}

    \tablecaption{sGRB Sample.  \label{tab:2}}

    \tablehead{\colhead{GRB} & \colhead{$T_{90}^{rest}$(s)} & \colhead{Redshift} & \colhead{$E_p$ (keV)} & \colhead{$L_p(\rm erg \, s^{-1})$ } & \colhead{References}}

   \startdata
    050709 &        0.6 &     0.16 &$      97.32 ^{+       7.76 }_{-       0.58 }$&$(     7.51 ^{+     0.76 }_{-     0.81 })\times 10^{50}$& (1)\\
    051221 &       0.91 &     0.55 &$     621.69 ^{+      87.42 }_{-     67.69 }$&$(       2.77 ^{+       0.29 }_{-       0.29 })\times 10^{52}$& (2)\\
    061006 &       0.35 &     0.44 &$     954.63 ^{+     198.39 }_{-    125.86 }$&$(       2.06 ^{+       0.15 }_{-       0.31 })\times 10^{52}$&(3) \\
    070714B &       1.04 &       0.92 &$     2150.4 ^{+     910.39 }_{-     443.52 }$&$(       6.56 ^{+       0.79 }_{-       1.36 })\times 10^{52}$&(4) \\
    090510 &       0.16 &      0.90 &$    8679.58 ^{+     947.69 }_{-     947.69 }$&$(        1.04 ^{+         0.24 }_{-       0.14 })\times 10^{54}$& (5)\\
    100117 &       0.16 &       0.92 &$     936.96 ^{+        297 }_{-        297 }$&$(       1.89 ^{+       0.21 }_{-       0.35 })\times 10^{52}$&(5)\\
    100206 &       0.09 &     0.41 &$     638.98 ^{+     131.21 }_{-     131.21 }$&$(      1.00 ^{+       1.15 }_{-     0.03 })\times 10^{52}$&(5) \\
    101219 &       0.35 &      0.72 &$     841.82 ^{+     107.56 }_{-       82.5 }$&$(       1.56 ^{+       0.24 }_{-       0.23 })\times 10^{52}$&(6) \\
    160821B &       0.41 &       0.16 &$      97.44 ^{+      22.04 }_{-      22.04 }$&$(    1.38 ^{+    0.18 }_{-    0.18 })\times 10^{50}$&(7) \\
    160624A &       0.13 &      0.48 &$     1247.2 ^{+      530.9 }_{-      530.9 }$&$(       1.85 ^{+      0.38}_{-      0.38 })\times 10^{52}$&(8) \\
    130603B &       0.13 &     0.36 &$     891.66 ^{+      135.1 }_{-      135.1 }$&$(       3.04 ^{+       0.44 }_{-       0.44 })\times 10^{52}$&(9) \\
    101224A &       0.12 &      0.72 &$     566.94 ^{+     690.63 }_{-      257.7 }$&$(      8.24 ^{+      1.25 }_{-      1.25 })\times 10^{51}$&(10)\\
    100625A &       0.23 &      0.45 &$     739.36 ^{+     112.53 }_{-     89.30 }$&$(        1.4 ^{+      0.06 }_{-      0.06 })\times 10^{52}$& (11)\\
    071227 &        1.30 &      0.38 &$       1000 ^{+   31.62 }_{-      32.62 }$&$(      4.43 ^{+      1.39 }_{-      1.29 })\times 10^{51}$&(12)\\
    070724A &       0.27 &      0.46 &$      56.49 ^{+      23.52 }_{-      13.16 }$&$(      8.7 ^{+       5 }_{-       5 })\times 10^{50}$& (13)\\
    050509B &       0.06 &      0.23 &$      74.15 ^{+     13.36 }_{-      50.31 }$&$(       1 ^{+      0.9 }_{-      0.9 })\times 10^{50}$&(14) \\
    \enddata

    \begin{flushleft}
    {Note. Each column corresponds to the name, the
    	rest-frame duration $T_{90}^{rest}$, the redshift, the peak energy $E_p$,
    	the peak luminosity $L_p$ in 64 ms of the observer frame time bin.} \\
    {Reference: (1) \citet{2005Natur.437..855V}; (2) \citet{2005GCN..4394....1G}; (3) \citet{2006GCN..5710....1G};
    (4) \citet{2007GCN..6638....1O}; (5) \citet{2013MNRAS.431.1398T}; (6) \citet{2010GCN.11470....1G};
    (7) \citet{2016GCN.19843....1S}; (8) \citet{2016GCN.19570....1H}; (9) \citet{2013GCN.14771....1G}; %
    (10) \citet{2010GCN.11489....1M}; (11) \citet{2010GCN.10912....1B}; (12) \citet{2007GCN..7155....1G};
    (13) \citet{2007GCNR...74....2Z}; (14) \citet{2005GCN..3385....1B}.}
\end{flushleft}
\end{deluxetable}

\newpage

\begin{deluxetable}{cccccccc}
    \tabletypesize{\footnotesize}
    \tablecaption{\textit{Fermi} sGRB Sample. \label{tab:fermi}}
    \tablehead{\colhead{name} & \colhead{$T_{90}$(s)}  & \colhead{z} & \colhead{flux(photons cm$^{-2}$ s$^{-1})$} & \colhead{$E_p$ (keV)} & \colhead{$\alpha$} & \colhead{$ \beta $} & \colhead{$ L( 10^{50} \rm erg \, s^{-1}) $}  }
    \startdata
GRB080723913&0.19&0.244&$4.88 \pm 0.97$&$124.99 \pm 133.73$&-1&-1.6&$14.7 \pm 2.65$ \\
GRB080725541&0.96&1.465&$6.93 \pm 1.11$&$470.5 \pm 333.61$&-0.89&-2.25&$388 \pm 98.6$ \\
GRB080802386&0.58&0.788&$10.9 \pm 1.52$&$370.65 \pm 123.85$&-0.42&-2.25&$156 \pm 31.5$ \\
GRB080815917&0.83&0.044&$6.49 \pm 1.52$&$22.71 \pm 42.21$&-1&-1.47&$0.684 \pm 0.153$ \\
GRB080831053&0.58&0.909&$4.21 \pm 0.84$&$162.81 \pm 63.37$&-1&-2.75&$45.4 \pm 15$ \\
GRB080905499&0.96&0.122&$6.5 \pm 0.88$&$676.74 \pm 274.76$&-0.17&-2.25&$2.69 \pm 0.431$ \\
GRB080919790&0.51&1.002&$4.89 \pm 0.99$&$161.92 \pm 107.49$&-0.7&-2.25&$48.6 \pm 33.6$ \\
GRB081012045&1.22&1.119&$6.86 \pm 0.95$&$1182.13 \pm 1020.84$&-0.47&-1.77&$1360 \pm 196$ \\
GRB081024245&0.83&1.785&$6.39 \pm 1.29$&$596.11 \pm 312.98$&-0.1&-4.98&$697 \pm 160$ \\
GRB081024891&0.64&2.419&$5.72 \pm 0.88$&$1495.9 \pm 1432.64$&-0.63&-2.11&$4360 \pm 670$ \\
GRB081101491&0.13&1.101&$7.06 \pm 1.07$&$207.75 \pm 90.31$&-0.46&-3.76&$79 \pm 23.2$ \\
GRB081102365&1.73&1.346&$5.66 \pm 1.05$&$463.68 \pm 450.11$&-0.63&-2.19&$350 \pm 92.4$ \\
GRB081105614&1.28&0.782&$6.8 \pm 0.78$&$324.88 \pm 98.75$&-1&-2.55&$125 \pm 17.8$ \\
GRB081107321&1.66&0.294&$13.03 \pm 1.31$&$58.5 \pm 13.03$&-1&-2.56&$4.54 \pm 0.767$ \\
GRB081119184&0.32&0.588&$4.38 \pm 2.2$&$57.26 \pm 14.17$&-1&-2.25&$6.12 \pm 3.34$ \\
GRB081122614&0.19&0.75&$8.23 \pm 1.35$&$176.52 \pm 72.46$&-0.62&-2.25&$44.9 \pm 12.1$ \\
GRB081204517&0.19&0.767&$11.4 \pm 1.19$&$363.41 \pm 116.03$&-0.44&-2.25&$148 \pm 30$ \\
GRB081209981&0.19&0.696&$25.54 \pm 1.65$&$758.37 \pm 149.16$&-0.2&-2.7&$460 \pm 32.5$ \\
GRB081216531&0.77&0.829&$37.66 \pm 1.71$&$1402.42 \pm 198.61$&-0.43&-3.71&$1420 \pm 69.5$ \\
GRB081223419&0.58&0.541&$14.36 \pm 1.2$&$199.67 \pm 37.6$&-0.36&-2.25&$44.7 \pm 5.81$ \\
GRB081226509&0.19&0.919&$7.91 \pm 0.94$&$284.87 \pm 88.83$&-1&-3.01&$114 \pm 19.9$ \\
GRB081229187&0.77&0.559&$4.32 \pm 0.79$&$188.7 \pm 106.63$&-1&-2.02&$41.5 \pm 8.51$ \\
GRB081230871&0.51&1.123&$3.97 $&$246.77 $&-1&-2.25&$106 $ \\
GRB090108020&0.7&0.353&$25.73 \pm 1.47$&$145.9 \pm 15.63$&-0.29&-2.25&$21.7 \pm 2.97$ \\
GRB090120627&1.86&6.883&$3.74 \pm 0.86$&$2541.31 \pm 1546.25$&-0.68&-2.25&$40400 \pm 8350$ \\
GRB090219074&0.45&0.098&$10.53 \pm 2.39$&$70.21 \pm 46.76$&-1&-1.63&$4.68 \pm 0.784$ \\
GRB090227772&0.3&0.373&$127.36 \pm 2.75$&$2029.8 \pm 92.38$&-0.09&-3.6&$1620 \pm 37.2$ \\
GRB090228204&0.45&0.33&$133.92 \pm 2.57$&$854.36 \pm 49.35$&-0.2&-3.39&$376 \pm 8.4$ \\
GRB090305052&1.86&0.68&$8.31 \pm 0.98$&$443.89 \pm 197.59$&-1&-1.96&$189 \pm 25.7$ \\
GRB090308734&1.66&1.272&$12.44 \pm 1.2$&$1035.2 \pm 501.14$&-0.63&-2.29&$1230 \pm 126$ \\
GRB090328713&0.19&1.029&$21.42 \pm 1.35$&$1203.96 \pm 525.45$&-0.82&-2.15&$1310 \pm 102$ \\
GRB090331681&0.83&1.622&$7.82 \pm 1$&$921.33 \pm 554.23$&-0.64&-2.25&$1280 \pm 186$ \\
GRB090405663&0.45&4.005&$7.28 \pm 1.09$&$1739.32 \pm 1995.18$&-0.85&-3.1&$10400 \pm 1660$ \\
GRB090412061&0.9&1.676&$3.94 \pm 0.95$&$576.93 \pm 441.21$&-0.25&-2.25&$619 \pm 195$ \\
GRB090429753&0.64&13.382&$3.49 \pm 0.22$&$1852.24 \pm 689.39$&-1.01&-2.25&$64300 \pm 4220$ \\
GRB090518080&2.05&1.407&$4.03 \pm 0.37$&$190 \pm 78.83$&-1.21&-2.25&$85.2 \pm 15.7$ \\
GRB090531775&0.77&5.818&$5.72 \pm 1$&$2695.72 \pm 2093.11$&-0.77&-2.76&$35100 \pm 5890$ \\
GRB090617208&0.19&0.484&$18.97 \pm 1.18$&$492.64 \pm 104.37$&-1&-2.18&$183 \pm 14.9$ \\
GRB090621922&0.38&0.315&$9.87 \pm 1.39$&$188.12 \pm 93.99$&-1&-1.84&$31.3 \pm 4.5$ \\
GRB090717111&0.38&1.592&$3.3 \pm 0.63$&$722.28 \pm 280.21$&-1&-2.25&$848 \pm 194$ \\
GRB090802235&0.13&0.4&$34.98 \pm 2.29$&$301.26 \pm 46.59$&-0.54&-2.25&$74.7 \pm 7.42$ \\
GRB090819607&0.19&0.837&$7.04 \pm 1.02$&$342.27 \pm 159.52$&-1&-2.24&$143 \pm 23.5$ \\
GRB090909854&1.15&0.205&$6.45 \pm 1.26$&$117.64 \pm 149.19$&-1&-1.6&$12.6 \pm 2.35$ \\
GRB090927422&0.51&0.625&$5.05 \pm 1.01$&$96.28 \pm 15.01$&-1&-2.25&$14.8 \pm 5.33$ \\
GRB091012783&0.7&0.708&$21.45 \pm 1.8$&$1049.71 \pm 223.79$&-0.25&-2.25&$790 \pm 62.1$ \\
GRB091018957&0.19&0.978&$9.25 \pm 1.81$&$340.37 \pm 153.33$&-0.11&-2.92&$160 \pm 30.8$ \\
GRB091019750&0.21&0.7&$7.64 \pm 0.95$&$164.27 \pm 59.18$&-1&-2.6&$38.1 \pm 9.75$ \\
GRB091126333&0.19&0.342&$10.16 \pm 1.24$&$173.56 \pm 81.56$&-1&-1.9&$28.4 \pm 3.72$ \\
GRB091223191&0.58&1.571&$4.27 $&$472.27 $&-0.36&-2.25&$418 $ \\
GRB091224373&0.77&0.067&$5.08 \pm 1.16$&$62.31 \pm 72.56$&-1&-1.34&$3.68 \pm 0.644$ \\
GRB100107074&0.58&0.161&$5.32 \pm 0.86$&$127.81 \pm 70.96$&-1&-1.55&$13.6 \pm 2$ \\
GRB100204858&1.92&1.725&$4.2 \pm 1.14$&$584.36 \pm 947.43$&-0.76&-1.99&$651 \pm 255$ \\
GRB100206563&0.18&0.407&$24.75 \pm 1.35$&$566.09 \pm 117.28$&-0.22&-2.38&$111 \pm 8.57$ \\
GRB100208386&0.19&1.369&$4.47 \pm 0.81$&$671.4 \pm 449.07$&-1&-2.14&$650 \pm 137$ \\
GRB100216422&0.19&0.668&$5.61 \pm 0.91$&$509.75 \pm 382.39$&-1&-1.71&$234 \pm 41.9$ \\
GRB100301068&0.96&0.988&$10.29 \pm 1.31$&$544.47 \pm 209.14$&-0.53&-2.25&$347 \pm 55$ \\
GRB100328141&0.38&0.692&$12.79 \pm 1.15$&$491.77 \pm 111.09$&-0.04&-2.25&$226 \pm 31.5$ \\
GRB100417166&0.19&1.071&$6.82 \pm 1.09$&$300.45 \pm 142.42$&-0.7&-2.25&$141 \pm 37.3$ \\
GRB100516396&0.64&1.061&$2.28 \pm 0.5$&$168.11 \pm 43.81$&-1&-3.16&$54.2 \pm 20.3$ \\
GRB100525744&1.47&0.265&$7.44 \pm 1.22$&$154.89 \pm 106.82$&-1&-1.71&$21.4 \pm 3.31$ \\
GRB100612545&0.58&0.612&$11.31 \pm 1.2$&$505.34 \pm 169.78$&-1&-2.09&$218 \pm 26.1$ \\
GRB100616773&0.19&0.958&$6.96 \pm 1.48$&$212.63 \pm 65.47$&-1&-4.53&$73.1 \pm 25.7$ \\
GRB100625773&0.24&0.452&$15.91 \pm 2.03$&$328.46 \pm 101.62$&-1&-2.23&$83.4 \pm 9.99$ \\
GRB100629801&0.83&0.493&$18.77 \pm 2.4$&$232 \pm 49.91$&-0.38&-2.25&$54.2 \pm 8.12$ \\
GRB100719311&1.6&2.675&$4 \pm 0.99$&$563.7 \pm 792.82$&-0.98&-2.25&$1000 \pm 404$ \\
GRB100805300&0.06&0.516&$14.32 \pm 2.69$&$205.76 \pm 46.5$&-1&-3.27&$45.7 \pm 7.28$ \\
GRB100811108&0.38&0.993&$18.56 \pm 1.18$&$998.67 \pm 186.92$&-1&-3.52&$937 \pm 67.5$ \\
GRB100816026&2.05&0.805&$19.39 \pm 1.4$&$131.92 \pm 23.64$&-0.13&-3.2&$81.6 \pm 12.5$ \\
GRB100827455&0.58&0.826&$19.92 \pm 1.2$&$814.17 \pm 169.36$&-0.21&-2.71&$582 \pm 42.5$ \\
GRB101021063&1.54&1.265&$4.32 \pm 1$&$839.35 \pm 1674.48$&-0.75&-1.62&$869 \pm 246$ \\
GRB101026034&0.26&0.169&$11.28 \pm 1.29$&$123.33 \pm 60.93$&-1&-1.66&$13 \pm 1.41$ \\
GRB101027230&1.34&1.064&$5.07 \pm 0.96$&$257.94 \pm 175.12$&-0.23&-2.4&$109 \pm 28$ \\
GRB101031625&0.38&0.798&$10.97 \pm 1.41$&$340.03 \pm 126.42$&-0.55&-2.25&$137 \pm 30$ \\
GRB101104810&1.28&1.383&$7.07 \pm 0.99$&$648.69 \pm 346.77$&-0.25&-2.54&$620 \pm 123$ \\
GRB101129652&0.38&1.091&$10.38 \pm 0.95$&$1261.81 \pm 346.77$&-0.14&-2.25&$1480 \pm 143$ \\
GRB101129726&0.58&1.541&$13.73 \pm 1.16$&$1266.12 \pm 650.81$&-0.84&-2.25&$2050 \pm 199$ \\
GRB101204343&0.13&0.611&$9.07 \pm 1.2$&$365.32 \pm 204.95$&-0.34&-1.95&$128 \pm 18.7$ \\
GRB101208203&0.19&0.199&$4.8 \pm 0.83$&$67.8 \pm 31.13$&-1&-1.79&$5.1 \pm 1.12$ \\
GRB101208498&2.05&0.394&$48.33 \pm 2.16$&$200.96 \pm 28.01$&-0.72&-4.36&$38.3 \pm 3.46$ \\
GRB101216721&1.92&0.582&$18.54 \pm 0.39$&$169.61 \pm 10.38$&-0.64&-3.47&$35.7 \pm 1.46$ \\
GRB110212550&0.06&0.675&$16.7 \pm 1.21$&$300.55 \pm 72.16$&-0.34&-2.76&$99.7 \pm 11.1$ \\
GRB110213876&0.32&0.557&$3.2 \pm 0.85$&$261.55 \pm 246.91$&-1&-1.72&$70.6 \pm 16.4$ \\
GRB110227009&1.73&0.968&$3.39 \pm 0.86$&$148.32 \pm 44.71$&-1&-2.25&$41 \pm 13.3$ \\
GRB110409179&0.13&0.559&$10.22 \pm 1.1$&$265.48 \pm 81.76$&-1&-2.23&$72.4 \pm 9.14$ \\
GRB110420946&0.13&0.965&$11.76 \pm 1.55$&$345.19 \pm 104.83$&-0.15&-4.33&$162 \pm 29.1$ \\
GRB110424758&0.67&0.232&$4.51 \pm 0.95$&$166.53 \pm 152.92$&-1&-1.52&$23.1 \pm 3.87$ \\
GRB110509475&0.64&0.107&$8.12 \pm 0.88$&$60.74 \pm 20.53$&-1&-1.65&$3.75 \pm 0.438$ \\
GRB110517453&0.58&1.232&$5.19 \pm 0.96$&$389.59 \pm 188.33$&-0.34&-2.25&$243 \pm 68.7$ \\
GRB110529034&0.51&1.195&$44.23 \pm 1.79$&$2314.86 \pm 548.33$&-0.9&-2.83&$4320 \pm 198$ \\
GRB110605780&1.54&0.978&$6.81 \pm 1.2$&$253.04 \pm 133.65$&-0.65&-2.25&$98.7 \pm 33.4$ \\
GRB110705151&0.19&0.64&$41.25 \pm 1.6$&$1007.63 \pm 127.59$&-0.13&-3.68&$692 \pm 29.9$ \\
GRB110717180&0.11&0.332&$19.11 \pm 1.63$&$260.98 \pm 82.03$&-1&-1.95&$54.5 \pm 4.96$ \\
GRB110728056&0.7&1.633&$4.97 \pm 0.9$&$786.04 \pm 431.17$&-0.27&-2.25&$998 \pm 172$ \\
GRB110801335&0.38&0.147&$11.76 \pm 2.8$&$54.94 \pm 38.26$&-1&-1.88&$3.37 \pm 0.611$ \\
GRB110916016&1.79&1.332&$3.26 \pm 0.96$&$263.68 \pm 134.02$&-1&-2.25&$138 \pm 60.8$ \\
GRB111001804&0.38&0.466&$3.82 \pm 0.81$&$170.63 \pm 116.48$&-1&-1.86&$31.9 \pm 6.88$ \\
GRB111011094&1.47&0.431&$15.28 \pm 1.27$&$202.65 \pm 56.69$&-1&-2.25&$40.6 \pm 4.81$ \\
GRB111022854&0.19&1.405&$7.08 \pm 1.11$&$471.19 \pm 352.04$&-0.64&-2.47&$374 \pm 91.5$ \\
GRB111024896&1.79&0.448&$5.01 \pm 0.99$&$232.12 \pm 161.53$&-1&-1.81&$51.6 \pm 10.1$ \\
GRB111103948&0.32&4.416&$7.88 \pm 1.07$&$2335.56 \pm 2125.41$&-0.91&-2.39&$19100 \pm 2580$ \\
GRB111112908&0.22&1.198&$15.23 \pm 1.12$&$896.52 \pm 213.2$&-0.28&-3.96&$921 \pm 69.4$ \\
GRB111117510&0.43&0.64&$10.09 \pm 0.97$&$284.65 \pm 71.12$&-1&-2.48&$88.1 \pm 10.1$ \\
GRB111207512&0.77&1.174&$3.16 \pm 0.81$&$225.08 \pm 79.77$&-1&-2.25&$95.1 \pm 45.4$ \\
GRB111222619&0.29&0.58&$64.25 \pm 3.29$&$787.53 \pm 121.24$&-0.68&-3.57&$436 \pm 20.9$ \\
GRB120101354&0.13&0.676&$8.39 \pm 1.04$&$185.16 \pm 48.75$&-0.11&-2.25&$45.3 \pm 17.2$ \\
GRB120129312&1.28&3.331&$2.73 \pm 0.72$&$1127.91 \pm 1135.85$&-0.33&-2.25&$4050 \pm 988$ \\
GRB120205285&0.58&0.2&$2.47 \pm 0.7$&$153.3 \pm 173.14$&-1&-1.39&$19.3 \pm 4.62$ \\
GRB120210650&1.34&3.353&$6.95 \pm 0.27$&$516.2 \pm 532.37$&-1.68&-3.15&$1140 \pm 101$ \\
GRB120212353&0.86&0.564&$5.83 \pm 1.37$&$362.32 \pm 859.93$&-0.77&-1.62&$121 \pm 29$ \\
GRB120222021&1.09&0.533&$22.73 \pm 1.72$&$174.25 \pm 33.97$&-0.55&-4.11&$35.5 \pm 6.91$ \\
GRB120519721&1.06&0.801&$14.27 \pm 1.5$&$379.96 \pm 140.77$&-0.32&-2.74&$164 \pm 23.2$ \\
GRB120524134&0.7&0.378&$16.71 \pm 1.82$&$89.99 \pm 18.27$&-0.51&-2.25&$10.2 \pm 6.78$ \\
GRB120603439&0.38&4.894&$2.65 \pm 0.24$&$882.49 \pm 418.03$&-0.81&-2.25&$4480 \pm 426$ \\
GRB120608489&0.96&1.84&$7.8 \pm 1.15$&$570.27 \pm 258.12$&-0.59&-4.91&$669 \pm 113$ \\
GRB120609580&1.79&6.873&$5.32 \pm 1.19$&$1053.73 \pm 3288.56$&-1.38&-2.25&$9600 \pm 2830$ \\
GRB120612687&0.26&0.445&$11.41 \pm 1.43$&$367.09 \pm 201.62$&-1&-1.83&$109 \pm 16.5$ \\
GRB120619884&0.96&5.077&$5.15 \pm 0.81$&$2972.37 \pm 2130.58$&-0.67&-2.31&$34100 \pm 5120$ \\
GRB120624309&0.64&0.883&$77.61 \pm 2.45$&$4100.76 \pm 445.32$&-0.71&-3.78&$8530 \pm 282$ \\
GRB120629565&0.7&0.639&$2.04 \pm 0.51$&$141.78 \pm 80.2$&-1&-2.04&$28.3 \pm 9.84$ \\
GRB120811014&0.45&0.615&$20.66 \pm 1.27$&$815.29 \pm 105.83$&-1&-2.25&$478 \pm 32.4$ \\
GRB120814201&0.9&3.644&$10.55 \pm 2.01$&$2293.79 \pm 3401.3$&-1.05&-2.06&$14400 \pm 2500$ \\
GRB120814803&0.19&1.069&$3.1 \pm 0.63$&$235.86 \pm 57.53$&-1&-2.25&$94.7 \pm 59.1$ \\
GRB120817168&0.16&0.58&$46.84 \pm 1.75$&$1068.72 \pm 183.42$&-0.66&-2.25&$716 \pm 32.7$ \\
GRB120830297&0.9&1.05&$10.28 \pm 0.86$&$919.7 \pm 218.87$&-1&-2.25&$857 \pm 80.7$ \\
GRB120831901&0.38&0.749&$9.73 \pm 1.43$&$320.69 \pm 108.24$&-0.19&-2.25&$119 \pm 28.5$ \\
GRB120915000&0.58&0.637&$5.15 \pm 1.07$&$508.98 \pm 547.59$&-0.23&-1.67&$227 \pm 50.4$ \\
GRB120916085&1.28&5.859&$4.57 \pm 1.01$&$1594.82 \pm 2252.8$&-0.95&-2.25&$15100 \pm 3280$ \\
GRB121004211&1.54&0.577&$5.02 \pm 2.13$&$74.24 \pm 11.65$&-1&-2.25&$9.25 \pm 11.8$ \\
GRB121012724&0.45&1.174&$11.64 \pm 1.12$&$650.85 \pm 195.38$&-0.26&-3.08&$537 \pm 68.2$ \\
GRB121023322&0.51&2.898&$10.86 \pm 1.29$&$1740.9 \pm 1882.78$&-1.02&-2.27&$6910 \pm 951$ \\
GRB121102064&2.05&3.097&$6.02 \pm 1.41$&$781.34 \pm 1440.09$&-1.18&-2.25&$2030 \pm 645$ \\
GRB121112806&1.28&1.44&$4.68 \pm 0.66$&$783.02 \pm 276.04$&-1&-2.25&$876 \pm 141$ \\
GRB121116459&0.83&1.034&$8.05 \pm 1.32$&$693.84 \pm 243.74$&-1&-2.25&$535 \pm 86.7$ \\
GRB121124606&0.26&1.512&$5.63 \pm 0.96$&$559.4 \pm 330.59$&-0.56&-2.25&$531 \pm 119$ \\
GRB121127914&0.64&0.731&$20.05 \pm 1.59$&$1014.58 \pm 205.6$&-0.26&-2.25&$764 \pm 56.1$ \\
GRB130112353&2.05&1.041&$8.49 \pm 1.05$&$374.43 \pm 160.5$&-0.72&-2.25&$197 \pm 40.1$ \\
GRB130127743&0.14&1.613&$8.06 \pm 0.94$&$1447.61 \pm 639.93$&-0.41&-2.25&$2670 \pm 333$ \\
GRB130204484&0.19&1.068&$9.62 \pm 1.26$&$653.13 \pm 403.8$&-0.51&-2.21&$498 \pm 83.4$ \\
GRB130219626&1.54&1.378&$6.78 \pm 1.09$&$334.66 \pm 133.24$&-0.24&-4.51&$210 \pm 48.5$ \\
GRB130307126&0.38&1.654&$17.47 \pm 1.28$&$1629.74 \pm 702.33$&-0.76&-2.88&$3320 \pm 261$ \\
GRB130325005&0.64&0.85&$6.37 \pm 0.97$&$166.58 \pm 68.42$&-0.62&-2.25&$44.8 \pm 18.7$ \\
GRB130404877&0.96&1.061&$4.31 \pm 1$&$216.81 \pm 92.19$&-0.11&-2.25&$82.1 \pm 29$ \\
GRB130416770&0.19&0.72&$24.06 \pm 1.34$&$1137.86 \pm 177.23$&-0.36&-2.25&$911 \pm 56.2$ \\
GRB130503214&0.88&0.535&$5.92 \pm 1.02$&$250.4 \pm 228.58$&-0.42&-1.8&$64.2 \pm 11.2$ \\
GRB130504314&0.38&0.809&$42.39 \pm 1.93$&$1397.06 \pm 180.06$&-0.47&-4.73&$1380 \pm 67$ \\
GRB130515056&0.26&0.539&$21.66 \pm 1.81$&$516.38 \pm 83.94$&-0.14&-2.25&$210 \pm 18.4$ \\
GRB130617564&0.77&1.223&$7.65 \pm 2.05$&$280.85 \pm 152.27$&-0.37&-4.92&$141 \pm 47.2$ \\
GRB130622615&0.96&1.272&$6.03 \pm 0.99$&$302.79 \pm 112.22$&-1&-4.87&$166 \pm 44.5$ \\
GRB130626452&1.73&0.991&$8.41 \pm 1.79$&$424.2 \pm 226.82$&-0.48&-2.25&$232 \pm 64.9$ \\
GRB130628860&0.51&0.656&$22.5 \pm 1.25$&$944.18 \pm 156.89$&-0.21&-2.25&$632 \pm 37.3$ \\
GRB130701761&1.6&0.765&$17.17 \pm 1.12$&$838.27 \pm 178.65$&-0.29&-2.25&$578 \pm 41.1$ \\
GRB130705398&0.13&1.077&$6.21 \pm 0.9$&$507.62 \pm 201.01$&-0.04&-2.25&$332 \pm 67.2$ \\
GRB130706900&0.13&1.1&$8.23 \pm 1$&$442.73 \pm 190.23$&-0.64&-2.25&$271 \pm 54.4$ \\
GRB130716442&0.77&2.213&$6.11 \pm 3.17$&$636.97 \pm 1427.42$&-0.54&-3.91&$980 \pm 1280$ \\
GRB130802730&0.06&0.689&$8.66 \pm 1.13$&$302.81 \pm 109.11$&-1&-2.32&$102 \pm 16.9$ \\
GRB130804023&0.96&0.388&$42.44 \pm 1.54$&$562.37 \pm 49.08$&-0.17&-2.25&$204 \pm 9.3$ \\
GRB130808253&0.26&0.35&$19.14 \pm 1.99$&$93.25 \pm 14.92$&-0.28&-2.25&$10.4 \pm 1.31$ \\
GRB130912358&0.51&1.597&$18.69 \pm 1.49$&$921.17 \pm 399.83$&-0.97&-4.86&$1260 \pm 115$ \\
GRB130919173&0.96&0.301&$16.93 \pm 1.28$&$86.71 \pm 20.06$&-1&-2.35&$8.7 \pm 1.09$ \\
GRB131004904&1.15&0.71&$9.66 \pm 1.6$&$143.98 \pm 89.53$&-1.25&-2.25&$31.9 \pm 9.33$ \\
GRB131126163&0.13&0.426&$34.4 \pm 1.87$&$451.03 \pm 71.42$&-1&-2.55&$148 \pm 10$ \\
GRB131128629&1.98&0.654&$5.46 \pm 1.08$&$86.63 \pm 28.23$&-0.02&-2.25&$12.9 \pm 32$ \\
GRB131217108&0.77&0.202&$7.84 \pm 1.07$&$223.54 \pm 113.55$&-1&-1.56&$35.8 \pm 4.3$ \\
GRB140105065&1.09&1.018&$8.64 \pm 1.08$&$420.6 \pm 173.15$&-0.59&-2.25&$233 \pm 46.6$ \\
GRB140105748&0.58&1.69&$4.43 \pm 1.04$&$440.58 \pm 347.76$&-0.67&-2.25&$402 \pm 133$ \\
GRB140109771&0.7&1.63&$5.66 \pm 1.63$&$725.92 \pm 517.15$&-0.48&-2.25&$875 \pm 243$ \\
GRB140129499&0.13&1.496&$7.38 \pm 1.24$&$589.58 \pm 364.13$&-0.8&-2.25&$572 \pm 115$ \\
GRB140209313&1.41&0.147&$123.96 \pm 3.26$&$164.36 \pm 14.28$&-0.06&-2.35&$20.1 \pm 0.689$ \\
GRB140329272&0.06&1.506&$7.82 \pm 0.91$&$754.49 \pm 362.25$&-0.68&-2.25&$862 \pm 120$ \\
GRB140402007&0.32&2.146&$5.27 \pm 0.77$&$1392.64 \pm 489.58$&-1&-2.25&$3390 \pm 500$ \\
GRB140428906&0.32&0.5&$17.9 \pm 2.02$&$490.04 \pm 220.62$&-0.21&-2.04&$185 \pm 24.8$ \\
GRB140501139&0.26&0.778&$5.09 \pm 1.04$&$473.8 \pm 383.6$&-0.04&-1.87&$231 \pm 47.2$ \\
GRB140511095&1.41&1.081&$11.02 \pm 1.29$&$315.47 \pm 107.25$&-0.57&-4.58&$154 \pm 29.2$ \\
GRB140518709&0.7&0.427&$4.08 \pm 0.79$&$182.84 \pm 69.28$&-1&-1.98&$34.1 \pm 6.4$ \\
GRB140526571&0.06&1.202&$4.33 \pm 0.83$&$421.25 \pm 157.46$&-1&-2.25&$270 \pm 74.5$ \\
GRB140605377&0.51&0.922&$11.19 \pm 0.9$&$728.34 \pm 170.85$&-1&-2.25&$528 \pm 49.9$ \\
GRB140610487&0.96&1.727&$4.86 \pm 0.97$&$1016.38 \pm 724.57$&-0.15&-2.25&$1610 \pm 391$ \\
GRB140619490&0.45&0.67&$11.77 \pm 2.09$&$307.91 \pm 77.37$&-0.23&-2.25&$103 \pm 17.6$ \\
GRB140624423&0.1&0.56&$20.42 \pm 1.47$&$292.08 \pm 55.38$&-0.71&-2.25&$84.7 \pm 8.93$ \\
GRB140626843&1.79&1.014&$8.78 \pm 1.69$&$246.93 \pm 199.88$&-0.75&-2.76&$97.7 \pm 30.4$ \\
GRB140710537&0.38&0.25&$7.42 \pm 1.26$&$248.43 \pm 245.41$&-0.03&-1.53&$45.4 \pm 7.98$ \\
GRB140720158&0.32&0.428&$6.12 \pm 1.07$&$306.97 \pm 252.95$&-1&-1.67&$79.5 \pm 13.8$ \\
GRB140724533&0.9&0.918&$4.46 \pm 0.91$&$184.45 \pm 125.53$&-1&-2.36&$56.1 \pm 19.9$ \\
GRB140807500&0.51&0.874&$20.88 \pm 1.42$&$826.85 \pm 234.32$&-0.74&-2.25&$623 \pm 48.9$ \\
GRB140901821&0.18&0.391&$52.28 \pm 1.81$&$697.62 \pm 62.93$&-0.31&-2.25&$290 \pm 11.5$ \\
GRB141011282&0.08&0.536&$30.07 \pm 1.54$&$475.68 \pm 98.47$&-0.46&-2.41&$183 \pm 13.1$ \\
GRB141031998&0.16&0.885&$4.69 \pm 0.84$&$304.34 \pm 190.86$&-1&-2.11&$123 \pm 25.8$ \\
GRB141102112&0.02&0.834&$8.25 \pm 1.4$&$265 \pm 107.56$&-0.45&-2.25&$94.1 \pm 26.8$ \\
GRB141105406&1.28&0.704&$10.27 \pm 1.01$&$322.96 \pm 53.57$&-1&-2.25&$115 \pm 15.2$ \\
GRB141111435&1.73&1.049&$2.62 \pm 0.69$&$628.63 \pm 732.78$&-1&-1.68&$461 \pm 121$ \\
GRB141113346&0.45&0.566&$4.49 \pm 1.06$&$234.88 \pm 217.77$&-1&-1.8&$59.7 \pm 12.1$ \\
GRB141122087&1.28&0.438&$6.27 \pm 1.1$&$285.95 \pm 234.62$&-1&-1.7&$71.6 \pm 12.3$ \\
GRB141124277&0.51&5.418&$4.9 \pm 1.47$&$1247.03 \pm 2306.81$&-1.09&-2.25&$9050 \pm 2630$ \\
GRB141128962&0.27&0.253&$12.99 \pm 1.38$&$58.14 \pm 30.19$&-0.34&-2.14&$4.27 \pm 0.809$ \\
GRB141202470&1.34&1.105&$11.13 \pm 0.99$&$647.6 \pm 140.91$&-1&-4.04&$505 \pm 56.8$ \\
GRB141205337&1.28&0.551&$5.29 \pm 1.3$&$362.55 \pm 359.96$&-1&-1.72&$119 \pm 25.8$ \\
GRB141208632&0.96&0.759&$7.09 \pm 1.26$&$465.5 \pm 572.62$&-0.76&-1.76&$220 \pm 45$ \\
GRB141213300&0.77&0.713&$15.09 \pm 1.36$&$195.26 \pm 66.57$&-0.75&-3.38&$51.2 \pm 10.3$ \\
GRB141230871&0.22&0.083&$7.78 \pm 1.25$&$40.8 \pm 34.51$&-1&-1.58&$1.89 \pm 0.278$ \\
GRB150101270&0.24&0.829&$5.73 \pm 1.07$&$213.22 \pm 125.07$&-0.45&-2.2&$65.7 \pm 16.5$ \\
GRB150101641&0.08&0.134&$10.06 \pm 1.42$&$23.1 \pm 4.75$&-1&-3.43&$0.27 \pm 0.054$ \\
GRB150118927&0.29&0.343&$39.78 \pm 1.81$&$257.32 \pm 42.98$&-0.39&-2.36&$53.9 \pm 3.13$ \\
GRB150128624&0.1&0.092&$15.93 \pm 3.44$&$64.91 \pm 68.7$&-1&-1.65&$4.08 \pm 0.612$ \\
GRB150208929&0.13&1.448&$5.95 \pm 0.88$&$788.16 \pm 404.28$&-0.28&-2.25&$891 \pm 145$ \\
GRB150214293&0.19&1.562&$5.75 \pm 1$&$462.92 \pm 289.02$&-0.44&-2.64&$403 \pm 93.7$ \\
GRB150301045&0.42&0.678&$8.56 \pm 1.61$&$97.49 \pm 47.49$&-0.8&-3.49&$15.9 \pm 6.2$ \\
GRB150312403&0.32&1.397&$6.58 \pm 1.03$&$485.6 \pm 274.43$&-0.73&-2.25&$391 \pm 87.2$ \\
GRB150316400&1.98&8.127&$2.66 \pm 0.88$&$910.24 \pm 2594.85$&-1.1&-2.25&$9620 \pm 4370$ \\
GRB150320462&0.06&0.838&$15.66 \pm 2.27$&$910.65 \pm 178.05$&-1&-2.25&$706 \pm 72$ \\
GRB150325696&0.08&0.84&$6.83 \pm 0.99$&$244.26 \pm 70.75$&-1&-2.25&$82.8 \pm 19.6$ \\
GRB150412507&0.58&0.901&$5.7 \pm 1.1$&$132.1 \pm 121.59$&-1.22&-2.25&$32.1 \pm 21.5$ \\
GRB150412931&0.64&2.281&$5.38 \pm 0.93$&$1142.19 \pm 790.59$&-0.57&-2.25&$2630 \pm 500$ \\
GRB150506630&0.51&1.355&$12.86 \pm 1.15$&$959.08 \pm 326.17$&-0.48&-2.91&$1150 \pm 104$ \\
GRB150506972&0.38&2.038&$4.32 \pm 0.92$&$642.36 \pm 589.55$&-0.61&-2.25&$907 \pm 257$ \\
GRB150522944&1.02&0.833&$4.12 \pm 0.89$&$265.86 \pm 256.51$&-0.17&-2&$94.5 \pm 22.3$ \\
GRB150601904&0.77&1.815&$2.41 \pm 0.51$&$573.57 \pm 220.21$&-1&-2.25&$666 \pm 177$ \\
GRB150604434&0.9&0.665&$11.23 \pm 1.11$&$348.06 \pm 115.88$&-0.03&-2.29&$125 \pm 14.7$ \\
GRB150605782&0.18&0.698&$7.78 \pm 1.34$&$152.22 \pm 49.01$&-0.39&-2.25&$33.6 \pm 7.66$ \\
GRB150609316&0.26&0.831&$5.23 \pm 0.95$&$185.59 \pm 48.68$&-1&-2.25&$52.5 \pm 22.8$ \\
GRB150628767&0.64&0.993&$3.57 \pm 0.69$&$218.64 \pm 46.67$&-1&-2.25&$78.7 \pm 35.4$ \\
GRB150629564&1.92&3.998&$9.18 \pm 1.01$&$1908.74 \pm 1178.45$&-0.92&-3.91&$12000 \pm 1390$ \\
GRB150705588&0.7&0.189&$6.96 \pm 1.28$&$55.25 \pm 53.82$&-1&-1.8&$3.6 \pm 0.808$ \\
GRB150715136&0.38&4.729&$4.41 \pm 0.84$&$1226.61 \pm 1100.27$&-0.71&-3.92&$7320 \pm 1370$ \\
GRB150721431&0.32&5.93&$5.13 \pm 1.04$&$2555.6 \pm 2163.57$&-0.8&-2.25&$33000 \pm 6370$ \\
GRB150728151&1.73&0.856&$7.04 \pm 1.09$&$574.74 \pm 425.92$&-0.23&-1.97&$339 \pm 70.7$ \\
GRB150805746&1.41&0.598&$5.91 \pm 1.18$&$58.39 \pm 21.37$&-0.01&-3.38&$6.38 \pm 2.22$ \\
GRB150810485&1.28&0.307&$17.14 \pm 1.13$&$242.62 \pm 52.37$&-1&-1.98&$46.9 \pm 3.24$ \\
GRB150811849&0.64&0.619&$25.83 \pm 1.83$&$745.9 \pm 151.85$&-0.43&-2.25&$415 \pm 33.1$ \\
GRB150819440&0.96&0.195&$146.49 \pm 3.43$&$542.06 \pm 26.1$&-0.05&-2.25&$150 \pm 4.61$ \\
GRB150828901&2.05&3.643&$1.59 \pm 0.22$&$492.97 \pm 1229.83$&-1.23&-1.7&$1180 \pm 245$ \\
GRB150901924&0.26&0.74&$5.25 \pm 1.08$&$225.34 \pm 32.04$&-1&-2.25&$66.3 \pm 17.6$ \\
GRB150906944&0.32&0.191&$15.34 \pm 1.52$&$157.16 \pm 68.89$&-1&-1.72&$19.9 \pm 1.82$ \\
GRB150912600&0.32&1.031&$6.03 \pm 0.95$&$382.28 \pm 142.27$&-0.12&-2.25&$202 \pm 53.5$ \\
GRB150922234&0.14&0.563&$24.07 \pm 1.29$&$675.11 \pm 95.15$&-0.19&-2.25&$333 \pm 19.8$ \\
GRB150923297&0.19&0.332&$8.06 \pm 1.02$&$189.92 \pm 170.63$&-0.47&-1.71&$32.5 \pm 4.42$ \\
GRB150923429&0.19&0.778&$5.52 \pm 0.85$&$227.18 \pm 41.42$&-1&-2.25&$69.6 \pm 14.1$ \\
GRB150923864&1.79&0.391&$16.45 \pm 1.19$&$120.89 \pm 21.12$&-1&-2.62&$16.7 \pm 2.2$ \\
GRB151022577&0.32&0.745&$11.46 \pm 1.53$&$322.58 \pm 117.41$&-0.48&-2.25&$120 \pm 28.1$ \\
GRB151202565&0.7&1.448&$8.06 \pm 1.67$&$660.37 \pm 499.08$&-0.75&-2.25&$667 \pm 165$ \\
GRB151222340&0.77&1.199&$13.25 \pm 1.78$&$1161.49 \pm 533.64$&-0.45&-2.43&$1410 \pm 167$ \\
GRB151228129&0.26&1.041&$12.32 \pm 1.73$&$963.99 \pm 351.43$&-0.4&-2.25&$919 \pm 128$ \\
GRB151229486&0.16&0.448&$5.56 \pm 1.25$&$226.93 \pm 216.49$&-1&-1.76&$49.7 \pm 10.8$ \\
GRB151231568&0.83&0.934&$14.27 \pm 1.2$&$578.08 \pm 237.7$&-0.7&-2.29&$366 \pm 43.8$ \\
GRB160211119&0.96&0.166&$2.66 \pm 0.97$&$65.97 \pm 163.56$&-1&-1.5&$4.66 \pm 1.4$ \\
GRB160224911&0.38&0.58&$4.09 \pm 1.09$&$132.52 \pm 108.92$&-1&-2.02&$23.8 \pm 7.35$ \\
GRB160314473&1.66&2.67&$6.4 \pm 1.12$&$709.24 \pm 989.35$&-1.18&-2.25&$1450 \pm 364$ \\
GRB160406503&0.43&0.26&$15.99 \pm 1.26$&$248.46 \pm 77.39$&-1&-1.79&$45.9 \pm 3.7$ \\
GRB160408268&1.06&0.73&$12.4 \pm 1.33$&$593.96 \pm 298.37$&-0.53&-1.99&$319 \pm 39.2$ \\
GRB160411062&0.67&0.659&$4.82 \pm 1.33$&$102.76 \pm 23.65$&-1&-2.25&$17.1 \pm 5.01$ \\
GRB160428412&0.58&7.849&$3.96 \pm 0.95$&$2133.7 \pm 4091.45$&-0.87&-2.11&$36700 \pm 8510$ \\
GRB160603719&0.38&1.218&$4.24 \pm 0.94$&$223.37 \pm 75.55$&-1&-4.3&$97.1 \pm 39$ \\
GRB160612842&0.38&0.483&$16.87 \pm 1.32$&$652.91 \pm 233.78$&-0.27&-1.86&$289 \pm 27.7$ \\
GRB160624477&0.45&0.483&$6.01 \pm 0.83$&$677.74 \pm 257.92$&-1&-2.25&$55.5 \pm 8.75$ \\
GRB160714097&0.32&0.212&$5.51 \pm 1.03$&$71.49 \pm 123.26$&-0.32&-1.7&$5.66 \pm 1.42$ \\
GRB160726065&0.77&0.998&$21.1 \pm 1.64$&$865.02 \pm 354.08$&-0.9&-2.25&$744 \pm 115$ \\
GRB160804180&0.64&0.957&$14.43 \pm 1.1$&$1198.61 \pm 347.69$&-0.3&-2.25&$1220 \pm 101$ \\
GRB160804968&0.32&1.918&$12.54 \pm 1.24$&$1722.47 \pm 920.84$&-0.82&-2.25&$4240 \pm 569$ \\
GRB160806584&1.73&0.503&$16.87 \pm 1.78$&$180.59 \pm 32.11$&-0.59&-2.25&$36.4 \pm 5.21$ \\
GRB160820496&0.38&1.358&$11.91 \pm 1.13$&$530.07 \pm 183.83$&-0.64&-4.88&$439 \pm 57.2$ \\
GRB160821937&1.09&0.16&$8.56 \pm 1.17$&$125.19 \pm 75.88$&-0.66&-2.41&$0.991 \pm 0.261$ \\
GRB160822672&0.42&0.626&$22.1 \pm 1.64$&$640.24 \pm 177.85$&-0.53&-2.16&$326 \pm 32.3$ \\
GRB160826938&1.79&0.446&$3.06 \pm 0.97$&$72.94 \pm 30.61$&-1&-2&$7.8 \pm 3.79$ \\
GRB160829334&0.51&0.546&$4.45 \pm 0.81$&$242.84 \pm 123.82$&-1&-1.94&$61.8 \pm 10.9$ \\
GRB161015400&0.19&0.457&$6.45 \pm 1.16$&$256.75 \pm 119.26$&-0.02&-1.8&$61.4 \pm 12.6$ \\
GRB161026373&0.11&1.15&$5.57 \pm 0.92$&$207.18 \pm 57.73$&-0.02&-3.56&$81.6 \pm 28.9$ \\
GRB161115745&0.03&1.642&$6.81 \pm 1.03$&$512.85 \pm 240.71$&-0.23&-4.11&$501 \pm 129$ \\
GRB161218222&0.32&1.152&$15.28 \pm 1.2$&$2041.9 \pm 377.69$&-0.4&-2.25&$3410 \pm 271$ \\
GRB161230298&0.45&0.949&$8.73 \pm 1.2$&$430.64 \pm 155.96$&-0.42&-2.25&$229 \pm 44.5$ \\
GRB170111760&0.83&1.178&$7.21 \pm 0.91$&$802.14 \pm 320.66$&-0.1&-2.25&$758 \pm 109$ \\
GRB170121133&1.6&0.556&$5.7 \pm 0.95$&$240.72 \pm 86.73$&-1&-2.1&$61.6 \pm 11.4$ \\
GRB170124528&0.45&0.445&$6.77 \pm 1.1$&$184.11 \pm 77.07$&-1&-2.08&$35.2 \pm 7.15$ \\
GRB170127067&0.21&0.373&$95.88 \pm 4.41$&$930.86 \pm 50.93$&-1&-4.96&$455 \pm 17.7$ \\
GRB170127634&1.73&0.847&$9.82 \pm 1.11$&$696.92 \pm 275$&-0.67&-1.85&$460 \pm 68$ \\
GRB170203486&0.34&0.184&$7.82 \pm 1.6$&$107.54 \pm 63.2$&-1&-1.64&$10.6 \pm 1.75$ \\
GRB170206453&1.17&0.353&$55.84 \pm 1.8$&$382.82 \pm 34.18$&-1&-2.77&$104 \pm 4.36$ \\
GRB170219002&0.1&1.038&$30.14 \pm 2.84$&$1292.15 \pm 256.4$&-0.62&-3.83&$1480 \pm 126$ \\
GRB170222209&1.66&1.33&$19.29 \pm 1.82$&$1346.51 \pm 367.49$&-0.89&-2.25&$1970 \pm 193$ \\
GRB170304003&0.16&0.342&$22.84 \pm 1.54$&$100.47 \pm 12.85$&-0.66&-2.25&$11.6 \pm 2.19$ \\
GRB170305256&0.45&0.443&$29.63 \pm 1.48$&$365.23 \pm 56.21$&-0.22&-2.38&$107 \pm 7.45$ \\
GRB170325331&0.58&2.278&$9.21 \pm 1.11$&$1658.22 \pm 1020.35$&-0.7&-2.48&$4820 \pm 649$ \\
GRB170506169&0.83&1.698&$9.16 \pm 1.49$&$1209.01 \pm 869.2$&-0.66&-2.25&$2100 \pm 446$ \\
GRB170511648&1.28&1.751&$3.67 \pm 0.9$&$185.98 \pm 83.75$&-0.93&-3.52&$102 \pm 65.7$ \\
GRB170604603&0.32&1.07&$12.5 \pm 1.13$&$1285.12 \pm 353.72$&-0.3&-2.25&$1500 \pm 158$ \\
GRB170817529&2.05&0.009&$2.94 \pm 0.76$&$176.97 \pm 98.81$&-1&-2.47&$0.00206 \pm 0.000774$ \\
    \enddata
\end{deluxetable}

\end{document}